\documentclass[prd,showpacs,showkeys,floatfix,nofootinbib,
               preprint,12pt,tightenlines,fleqn]{revtex4} %for preprint

\usepackage{amsmath,amssymb,revsymb,graphicx,dcolumn}
\newcommand{\beq}{\begin{equation}}
\newcommand{\eeq}{\end{equation}}
\newcommand{\beqa}{\begin{eqnarray}}
\newcommand{\eeqa}{\end{eqnarray}}
\newcommand{\bsubeqs}{\begin{subequations}}
\newcommand{\esubeqs}{\end{subequations}}
\begin{document}
\noindent Phys. Rev. D 86, 027302 (2012)
\hfill arXiv:1204.5085\vspace*{8mm}\newline
%\noindent arXiv:1204.5085 \hfill KA--TP--16--2012\;(\version)\vspace*{8mm}\newline
\title[De Sitter-spacetime instability]
      {De Sitter-spacetime instability from a nonstandard vector field\vspace*{5mm}}
\author{V. Emelyanov}
\email{slawa@particle.uni-karlsruhe.de}
\author{F.R. Klinkhamer}
\email{frans.klinkhamer@kit.edu}\affiliation{Institute for
Theoretical Physics,\\
Karlsruhe Institute of Technology (KIT),\\
76128 Karlsruhe, Germany\\}

\begin{abstract}
\vspace*{2.5mm}\noindent
It is found that de Sitter spacetime, the constant-curvature matter-free
solution of the Einstein equations with a positive cosmological constant,
becomes classically unstable due to the dynamic effects of a certain
type of vector field (fundamentally different from a gauge field).
The perturbed de Sitter universe evolves towards a
final singularity. The relevant vector-field configurations
violate the strong and dominant energy conditions.
\end{abstract}



\maketitle

\section{Introduction}
\label{sec:Introduction}

The Einstein gravitational field equations with positive
cosmological constant~\cite{Einstein1917}
have a highly symmetric matter-free solution,
de Sitter spacetime~\cite{deSitter1917,HawkingEllis1973}.
Nearly a century after the discovery of this mathematical solution,
de Sitter spacetime occupies a central place in modern theoretical physics
and observational cosmology
(see, for example, the reviews~\cite{Weinberg1989,PDG2010}).
It is, then, all the more interesting if something new can be
said about de Sitter spacetime, even if the context is nonstandard.

In recent work on the cosmological constant problem, we noted parenthetically
(Footnote~1 in Appendix~A of Ref.~\cite{EmelyanovKlinkhamer2011-CCP1-FRW-NEWTON})
that, for the simple model considered,
de Sitter spacetime corresponded to an \emph{unstable} critical point.
The simple model
considered~\cite{Dolgov1997,EmelyanovKlinkhamer2011-CCP1-NEWTON}
had a classical vector field $V_\alpha(x)$ with a ``wrong-sign''
kinetic term [giving energy density $\rho_\text{vec}\leq 0$
for the cosmological solution], which we suspected
to be responsible for the de Sitter instability.
It will, however, be shown in the present article that
the de Sitter instability is also present in the model
with a ``correct-sign'' kinetic term
[giving $\rho_\text{vec}\geq 0$ for the cosmological solution].

The particular type of vector-field theory
considered (Sec.~\ref{sec:Theory}) is, most likely,
pathological, having instabilities at the classical
level and ghosts at the quantum level.
Still, the vector field interacts only
gravitationally with the other matter fields. As such, this classical vector
field may be used to describe certain nonstandard
gravitational effects in the long-wavelength (low-energy) limit. Two examples of such
effects are discussed in the present article, namely,
a particular type of instability of the
de Sitter equilibrium solution (Sec.~\ref{sec:Unstable-equilibrium})
and the corresponding
final singularity (Sec.~\ref{sec:Final-singularity}). The estimated de Sitter
decay time and the violation of certain energy conditions by the relevant
vector-field configurations are discussed in Sec.~\ref{sec:Discussion}. In
that last section, it is also explained how this type of classical vector
field can perhaps play a role in the macroscopic description
of a fundamental quantum instability of de Sitter
spacetime~\cite{Polyakov2007,Polyakov2009,KrotovPolyakov2010}.

%%\newpage%%tmp
\section{Theory}
\label{sec:Theory}

Consider general relativity with a positive
cosmological constant $\Lambda$ and
a single classical vector field $V_{\alpha}(x)$.
The specific gravitational
model~\cite{Dolgov1997,EmelyanovKlinkhamer2011-CCP1-FRW-NEWTON,%
EmelyanovKlinkhamer2011-CCP1-NEWTON} used here
has the following action \mbox{($c=\hbar=1$):}%
\bsubeqs\label{eq:model-action-Q1-EPlanck}
\beqa\label{eq:model-action}
\hspace*{-10mm}
S[g,\, V,\, \phi]
&=&
-\int\,d^4x\,\sqrt{-\text{det}(g)}\,
\bigg( \frac{1}{2}\,(E_\text{Planck})^{2}\,R[g]
+ \epsilon\big(Q_{1}[g,\, V]\big)
+ \Lambda
+ \mathcal{L}_{M}[g,\,\phi]\bigg),
\\[2mm]
\label{eq:epsilonQ1}
\hspace*{-10mm}
\epsilon\big(Q_{1}[g,\, V]\big)
&=&
-\big(Q_{1}[g,\, V]\big)^{2}\equiv -V_{\alpha;\beta}\,V^{\alpha;\beta}\,,
\\[2mm]\label{eq:EPlanck}
\hspace*{-10mm}
E_\text{Planck}
&\equiv&
(8\pi G)^{-1/2}\,,\quad G >0\,,\quad \Lambda >0\,,
\eeqa
\esubeqs
where
a generic massless matter field $\phi(x)$ has been added with
a standard Lagrange density $\mathcal{L}_{M}(x)$.
The action \eqref{eq:model-action} is really classical, but,
for convenience, we use quantum terminology such as $E_\text{Planck}$.
In principle, it is also possible to add
a mass term for the vector field, but we refrain from doing so for
the moment and the theory maintains the shift invariance
of the vector field.

Notice that, unlike the case of a gauge field
with a Maxwell action-density term, the time derivative
of the $V_0$ component enters the action-density
term  \eqref{eq:epsilonQ1}. It is, of course, known that,
in Minkowski spacetime ($\Lambda=0$),
gauge invariance is required for the
Poincar\'{e} invariance, locality, and stability of the
massless-vector-field theory~\cite{Weinberg1964}.
However, as explained in Sec.~\ref{sec:Introduction}, our interest in the
classical massless vector field from \eqref{eq:model-action-Q1-EPlanck}
is only as an effective way to describe possible nonstandard gravitational
effects related to the cosmological constant $\Lambda>0$.
Our focus will be on stability issues in a cosmological context.

Let us restrict our attention to the spatially flat ($k=0$)
Robertson--Walker metric~\cite{HawkingEllis1973}
with a perfect-fluid standard-matter component
and an isotropic vector field (vanishing spatial components
in appropriate coordinates). The dimensionless
ordinary differential equations (ODEs) are then
\bsubeqs\label{eq:ODEs}
\beqa\label{eq:ODEs-friedmann}
\hspace*{-4mm}
&&
3\,h^{2} = 1 + \dot{v}^{2} + 3\,h^{2}\,v^{2}+\lambda^{-1}\,r_{M}
\,\,,
\\[2mm]\label{eq:ODEs-hdot}
\hspace*{-4mm}
&&
2\,\dot{h} = 2\,\dot{h}\,v^{2} + 4\,h\,v\,\dot{v} - 2\,\dot{v}^{2}
-\lambda^{-1}\,(1+w_{M})\,r_{M}\,,
\\[2mm]\label{eq:ODEs-vdot}
\hspace*{-4mm}
&&
\ddot{v} + 3\,h\,\dot{v} - 3\,h^{2}\,v
= 0\,,
\\[2mm]\label{eq:ODEs-rMdot}
\hspace*{-4mm}
&&
\dot{r}_{M} + 3\,(1+w_{M})\,h\,r_{M} =0\,,
\eeqa
\esubeqs
where a numerical factor
$\sqrt{\lambda} \equiv \sqrt{\Lambda}/(E_\text{Planck})^{2}>0$
has been absorbed into the definitions of
the dimensionless  inverse Hubble parameter $h^{-1}$ and
the dimensionless cosmic time $\tau$
(the overdot stands for differentiation with respect to this $\tau$).
The dimensionless variable $v$ corresponds to
the vector-field time-component $V_{0}$
and the dimensionless variable $r_{M}\geq 0$ corresponds
to the standard-matter energy density $\rho_{M}\geq 0$ with constant
equation-of-state parameter $w_{M}\geq 0$.
See Appendix~A of Ref.~\cite{EmelyanovKlinkhamer2011-CCP1-FRW-NEWTON}
for further details.

%%\newpage%%tmp
Using Eq.~(A6) from Ref.~\cite{EmelyanovKlinkhamer2011-CCP1-NEWTON},
the corresponding dimensionless vector-field energy density $r_\text{vec}$
and pressure $p_\text{vec}$ are found to be given by
\bsubeqs\label{eq:rvec-pvec}
\beqa\label{eq:rvec}
r_\text{vec} &\equiv&
e(q_{1}) - q_{1}\, \frac{d e(q_{1})}{d q_{1}}
=
+(q_{1})^{2}= \lambda\,(\dot{v}^{2} + 3\,h^{2}\,v^{2}) \geq 0\,,
\\[2mm]\label{eq:pvec}
p_\text{vec}&=&-r_\text{vec}-2\,\lambda\,\big(\dot{h}\,v^{2}+2\,h\,v\,\dot{v}-\dot{v}^{2}\big)\,,
\eeqa
\esubeqs
parts of which, divided by $\lambda$,
can be seen to appear on the right-hand sides
of \eqref{eq:ODEs-friedmann} and \eqref{eq:ODEs-hdot}.
Remark that, in a Minkowski background with $H=\Lambda=0$,
the vector-field fluid \eqref{eq:rvec-pvec} is not unusual,
it corresponds to a matter component with an ultrahard equation of state,
$\rho_\text{vec}=P_\text{vec}=(dV_{0}/dt)^2 \geq 0$.
What is unusual is how the vector-field pressure \eqref{eq:pvec}
behaves in a nonflat spacetime background,
possibly having $P_\text{vec} < -\rho_\text{vec}<0$.

%%\newpage%%tmp
\section{Unstable equilibrium}
\label{sec:Unstable-equilibrium}

The ODEs \eqref{eq:ODEs} have an asymptotic
equilibrium solution (critical point) corresponding to de Sitter spacetime:
\bsubeqs\label{eq:deS-critical point}
\beqa
1/h(\tau)&=&\sqrt{3}\,,
\\[2mm]
v(\tau)&=&0\,,
\\[2mm]
r_{M}(\tau)&=&0\,.
\eeqa
\esubeqs
There are no asymptotic solutions with $h \sim 0$,
which would approach
Minkowski spacetime~\cite{EmelyanovKlinkhamer2011-CCP1-FRW-NEWTON,%
Dolgov1997,EmelyanovKlinkhamer2011-CCP1-NEWTON}. The explanation is that,
with the minus sign chosen in \eqref{eq:epsilonQ1},
the cosmological constant cannot be canceled by the
vector-field contribution, as the right-hand side of
\eqref{eq:ODEs-friedmann} makes clear
($\lambda^{-1}\,r_{M}$ is non-negative).

The equilibrium solution \eqref{eq:deS-critical point}
is, however, unstable.
We will show this numerically, but it can also be proven
mathematically by following the discussion in
Appendix~A of Ref.~\cite{EmelyanovKlinkhamer2011-CCP1-FRW-NEWTON}
(in fact, the linearized analysis suffices, according to
Theorem 3.2 of Ref.~\cite{Verhulst1996}).

\begin{figure*}[t]  %%frk: choose [p] or [t]   %%frk
\begin{center}                         %%fig1_v5801.eps = fig1_v6.eps
\hspace*{-7mm}
\includegraphics[width=1.06\textwidth]{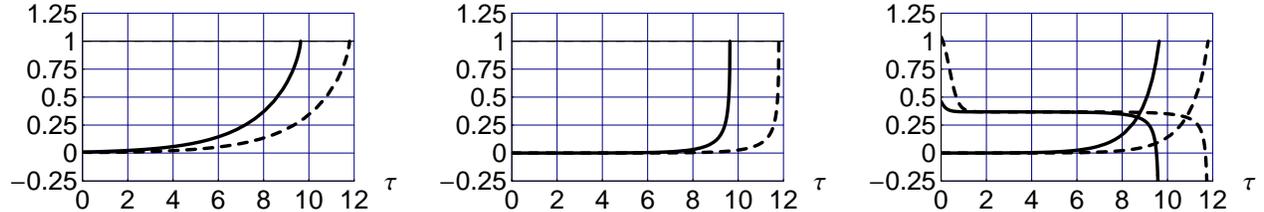}
\end{center}
\vspace*{-8mm} \caption{Numerical solutions of ODEs \eqref{eq:ODEs-hdot} and
\eqref{eq:ODEs-vdot} with dimensionless cosmological constant $\lambda>0$ and
vanishing standard-matter component, $r_{M}(\tau)=0$.
The left panel shows the dimensionless vector-field component $v(\tau)$.
The middle panel shows $(3\,h^{2})^{-1}\,\dot{v}^{2}/(1-v^{2})$.
The right panel shows
$r_\text{vec}/(3\,\lambda\,h^{2})$ as the ascending curves
and
$3/4+(1/4)\,(p_\text{vec}/r_\text{vec})$ as the descending curves.
The vector-field
boundary conditions are $v(0)=1/130$ and $\dot{v}(0)=\pm 1/130$, where the
solid and dashed curves correspond to the plus and minus sign, respectively.
The values for $h(0)$ follow from \eqref{eq:ODEs-friedmann} [both curves
have, in fact, the same value of $h(0)$].
The Hubble parameter $h(\tau)$ and
the vector-field energy density $r_\text{vec}(\tau)$ diverge
at $\tau \approx 9.635$ for the $\dot{v}(0)>0$ boundary condition
(ascending solid curve of the right panel)
and at $\tau \approx 11.794$ for the $\dot{v}(0)<0$ boundary condition
(ascending dashed curve of the right panel).\vspace*{0cm}} \label{fig:01}
\end{figure*}

Numerical solutions have been obtained
with vanishing and nonvanishing standard-matter components.
For the case of a nonvanishing standard-matter component,
it is found that an asymptotic de Sitter spacetime is approached
if the vector field is strictly equal to zero, but not
if the vector field is nonzero.
As the conclusion is essentially the same
for the case of a vanishing standard-matter component,
we focus on the $r_{M}=0$ case~\cite{endnote:rMnonzero}.
Instead of reaching an asymptotic de Sitter spacetime,
the model universe of Fig.~\ref{fig:01} is seen to terminate after
a finite time interval, having a diverging
Ricci scalar $R \propto -6\,(\dot{h}+2\,h^{2})$ at $\tau \sim 10$.
The same type of behavior as shown in Fig.~\ref{fig:01} is obtained
for boundary conditions taking values in the finite intervals
$v(0)\in (0,\,1/100]$ and $\dot{v}(0)\in [-1/100,\,+1/100]$.

Hence, we have established numerically
the classical instability of de Sitter
spacetime \eqref{eq:deS-critical point} under vector-field
perturbations $\delta v(\tau)$ which break the original
de Sitter symmetry. We have, in addition,
explicit analytic results
for the linear perturbations but will not present them here,
as the numerical results suffice to demonstrate the instability.

%%\newpage%%tmp
\section{Final singularity}
\label{sec:Final-singularity}

The numerical results of the previous section suggest
that the model universe of Fig.~\ref{fig:01}
runs into a final singularity (also known as a big-rip-type future
singularity or, more generally, as an exotic future singularity,
see Refs.~\cite{Caldwell1999,Starobinsky1999,McInnes2001,%
Caldwell-etal2003,Faraoni2003,Nojiri-etal2005,Dabrowski2011,Barrow-etal2012}
and references therein).
Some analytic results have been obtained for the
vector-field theory \eqref{eq:model-action-Q1-EPlanck}
with vanishing standard-matter component, $r_{M}(\tau)=0$.
The pure vector-field theory considered here is of interest, because
it has a strictly non-negative energy density \eqref{eq:rvec},
different from the scalar-field theory with a negative quartic
coupling constant as discussed in Ref.~\cite{Faraoni2003}.

The ODEs \eqref{eq:ODEs} now reduce to
\bsubeqs\label{eq:ODEs-rMzero}
\beqa
\label{eq:ODEs-rMzero-vdot}
\ddot{v} + 3\,h\,\dot{v} - 3\,h^{2}\,v&=& 0\,,
\\[2mm]
\label{eq:ODEs-rMzero-hdot}
\big(\dot{h}\big)^{-1}
- \left(\frac{2\,h\,v\,\dot{v}-\dot{v}^{2}}{1-v^{2}}\right)^{-1}&=&0\,,
\\[2mm]
\label{eq:ODEs-rMzero-friedmann}
\frac{h^{-2} + h^{-2}\,\dot{v}^{2}}{3\,(1-v^{2})}&=&1\,\,,
\eeqa
\esubeqs
where $h$, $\dot{h}$, and $1-v^{2}$ are assumed to be nonzero.
Next, make a change of variable $s = \ln(a)$
for cosmic scale factor $a(\tau)$, with $a(0)=1$
and Hubble parameter $h\equiv \dot{a}/a \equiv d s/d \tau$.
The following ODEs for $v(s)$ and $h(s)$ are found:
\bsubeqs\label{eq:ODEs-rMzero-vdoubleprime-hprime}
\beqa\label{eq:ODEs-rMzero-vprimeprime}
v'' + \frac{2\,v-v'}{1-v^{2}}\,(v')^{2}+3\,(v'-v)&=&0\,,
\\[2mm]\label{eq:ODEs-rMzero-hprime}
\big(h^{\prime}\big)^{-1}
- \left(\frac{2\,v - v^{\prime}}{1-v^{2}}\,v^{\prime}\,h\right)^{-1} &=&0\,,
\\[2mm]
\label{eq:ODEs-rMzero-friedmann-prime}
\frac{h^{-2} + (v')^{2}}{3\,(1-v^{2})}&=&1\,\,,
\eeqa
\esubeqs
where the prime stands for differentiation with respect to $s$.
There are only two arbitrary constants of integration
as the last two ODEs in \eqref{eq:ODEs-rMzero-vdoubleprime-hprime}
are first-order (the first ODE is consistent with the last two
ODEs; cf. Sec.~III A of Ref.~\cite{EmelyanovKlinkhamer2011-CCP1-FRW-NEWTON}).

It is easy to check that a particular combination of
trigonometric functions solves the nonlinear
ODE \eqref{eq:ODEs-rMzero-vprimeprime}.
With an arbitrary real amplitude $A\in [-1,\,+1]$
and an arbitrary relative sign entering the
solution for $v(s)$
and with a nonzero real amplitude $B$ in $h(s)$,
the following solutions of the ODEs \eqref{eq:ODEs-rMzero-vprimeprime}
and \eqref{eq:ODEs-rMzero-hprime} are obtained:
\bsubeqs\label{eq:general-critical-point}
\beqa
\label{eq:general-critical-point-v}
%\hspace*{-12mm}
v(s)&=& A\,\sin\big(\sqrt{3}\,s\big)
\pm \sqrt{1-A^{2}}\,\cos\big(\sqrt{3}\,s\big)\,,
\\[2mm]
\label{eq:general-critical-point-h-inverse}
%\hspace*{-12mm}
1/h(s) &=&
B\,\big[1 + (2A^{2} - 1)\cos(2\sqrt{3}\,s)
\mp 2A\,\sqrt{1-A^{2}}\,\sin(2\sqrt{3}\,s)\big]\,\exp(3\,s)\,.
\eeqa
\esubeqs

A consistent solution of the
differential system \eqref{eq:ODEs-rMzero-vdoubleprime-hprime} with
$v$ and $h$ given by \eqref{eq:general-critical-point-v}
and \eqref{eq:general-critical-point-h-inverse}
requires a solution of the
constraint~\eqref{eq:ODEs-rMzero-friedmann-prime}.
From \eqref{eq:general-critical-point-v},
this implies $h^{-2}=0$, which is only possible for special values of $s$
according to \eqref{eq:general-critical-point-h-inverse}.
It turns out that $|v|=1$ for the values of $s$ that nullify
$1/h$. One concrete example
has $A=1$ and $s=\pi/(2\sqrt{3})$, while keeping an
arbitrary nonzero $B$.
Specifically, we have for this particular
critical point of the differential system \eqref{eq:ODEs-rMzero-vdoubleprime-hprime}:%
\bsubeqs\label{eq:final-critical-point}
\beqa
\label{eq:final-critical-point-h-inverse}
\Big[1/h(s)\,\Big]_{s=\pi/(2\sqrt{3})} &=&
\Big[B\,\big[1 + \cos(2\sqrt{3}\,s)\big]\,e^{3\,s}\,\Big]_{s=\pi/(2\sqrt{3})}
%%\nonumber\\[1mm]&=&0\,,   %%frk
=0\,,
\\[2mm]
\label{eq:final-critical-point-v}
\Big[v(s)\,\Big]_{s=\pi/(2\sqrt{3})}&=&
\Big[\sin\big(\sqrt{3}\,s\big)\,\Big]_{s=\pi/(2\sqrt{3})}=1\,.
\eeqa
\esubeqs
The actual value for $s$ at the critical point is nonphysical
(because $a$ is);
what matters is that, for example, the Ricci scalar diverges there.
For cosmic times just before the singularity, the functions
$v$ and $1/h$ can be expected to be slightly different
from those given in \eqref{eq:general-critical-point}.

Moreover, the following corollary can be obtained from
\eqref{eq:ODEs-rMzero-friedmann-prime}:
\bsubeqs\label{eq:final-critical-point-ratios}
\beq\label{eq:final-critical-point-ratio-v}
\left[
\frac{1}{3}\;\frac{(v^\prime)^{2}}{1-v^{2}}\,\right]_\text{singularity}=1\,,
\eeq
where the suffix is interpreted as being arbitrarily close
to the point with $1/h(s) = d \tau/d s=0$
corresponding to \eqref{eq:final-critical-point-h-inverse}.
The explicit solution \eqref{eq:general-critical-point-v}
can also be seen to satisfy \eqref{eq:final-critical-point-ratio-v}.
In turn, \eqref{eq:final-critical-point-ratio-v} gives
for the vector-field energy density \eqref{eq:rvec}
the following result:
\beq\label{eq:final-critical-point-ratio-rvec}
\left[\frac{r_\text{vec}}{3\,\lambda\,h^{2}}\,\right]_\text{singularity}=
\left[\frac{\rho_\text{vec}}{3\,(E_\text{Planck})^{2}\,H^{2}}\,\right]_\text{singularity}=1\,,
\eeq
with dimensional quantities in the middle expression.
A final characteristic concerns the diverging vector-field
equation-of-state parameter $w_\text{vec}\equiv p_\text{vec}/r_\text{vec}$
and can be stated as follows:
\beq\label{eq:final-critical-point-ratio-pvec-over=rvec}
\left[
\sqrt{3\,(1-v^2)/16}\;\;
\frac{r_\text{vec}+p_\text{vec}}{r_\text{vec}}\,\right]_\text{singularity}
= -1 \,,
\eeq
\esubeqs
which, using \eqref{eq:final-critical-point-ratio-rvec},
can also be written with $3\,\lambda\,h^{2}$ in the denominator.

Further mathematical discussion of the final
singularity \eqref{eq:final-critical-point}
is left to a future publication.
Note that, strictly speaking, the qualification ``final'' is arbitrary,
as the tensor-vector-scalar theory \eqref{eq:model-action-Q1-EPlanck}
is time-reversal-invariant
and so is the differential system \eqref{eq:ODEs-rMzero}.

For the present article, the relevant observation is
that the numerical results of Fig.~\ref{fig:01} can be interpreted
as interpolating between the critical points
\eqref{eq:deS-critical point} and \eqref{eq:final-critical-point}.
Indeed, the particular combination $3^{-1}\,h^{-2}\,(\dot{v})^{2}/(1-v^{2})$
from the middle panel of Fig.~\ref{fig:01}
is seen to run between the values $0$ and $1$,
which are the corresponding values
from \eqref{eq:deS-critical point}
and \eqref{eq:final-critical-point-ratio-v}.
The numerical results for the ratio $r_\text{vec}/(3\,\lambda\,h^{2})$
from the right panel of Fig.~\ref{fig:01}
show the same behavior, running between the values $0$ and $1$
from \eqref{eq:deS-critical point}
and \eqref{eq:final-critical-point-ratio-rvec}, respectively.
Numerical results also match \eqref{eq:final-critical-point-ratio-pvec-over=rvec},
but have not been shown explicitly in Fig.~\ref{fig:01}.

Including standard-matter, the final singularity
is characterized by having $\rho_\text{vec}/\Lambda\to\infty$ and
$\rho_{M}/\Lambda\to \text{const} >0$,
in addition to having a diverging Ricci scalar $R$ as mentioned before.
This exotic behavior, just as that of Fig.~\ref{fig:01} for
the $\rho_{M}=0$ case,
is, most likely, the result of the unusual
properties of the vector-field energy density and
pressure, which will be discussed in the next section.

%%\newpage%%tmp
\section{Discussion}
\label{sec:Discussion}

Let us, first,
elaborate on the remark of Sec.~\ref{sec:Unstable-equilibrium}
about the finite age of
the type of model universe shown in Fig.~\ref{fig:01}.
For initial values of the standard-matter energy density
that are not too large (compared to the value of
$\Lambda$), one has an age of the order of
\bsubeqs\label{eq:tmax-cin}
\beq\label{eq:tmax}
t_\text{max} - t_\text{in}
\sim c_\text{in}\;E_\text{Planck}/\sqrt{\Lambda}\,,
\eeq
where the numerical coefficient
$c_\text{in}$ depends on the
vector-field boundary conditions
at $t=t_\text{in}$
($c_\text{in}\sim 10$ for the boundary conditions of Fig.~\ref{fig:01}).
From the analytic solution of the vector-field equation
\eqref{eq:ODEs-vdot}
with $h$ replaced by $1/\sqrt{3}$, it is estimated that
the dependence of $c_\text{in}$ on
the initial values is only logarithmic,
\beqa\label{eq:cin}
c_\text{in}
&\sim& c_1\,
\ln\left(\frac{1}{|\,c_2\,v_\text{in} + c_3\,\dot{v}_\text{in}|}\right)
\,,
\\[2mm]\label{eq:vin}
v_\text{in}
&\equiv& (E_\text{Planck})^{-1}\;V_0(t_\text{in})\,,
\\[2mm]\label{eq:vdotin}
\dot{v}_\text{in}
&\equiv& \Lambda^{-1/2}\;
\left.\frac{d V_0(t)}{d t}\right|_{t=t_\text{in}}
\,,
\eeqa
\esubeqs
with positive constants $c_1,\,c_2,\,c_3$ of order 1
and generic small values \eqref{eq:vin} and \eqref{eq:vdotin},
making for a positive and finite logarithm in \eqref{eq:cin}.

In a de Sitter spacetime with Hubble constant $H_\text{dS}$,
the Gibbons--Hawking
temperature $T_\text{GH}$~\cite{GibbonsHawking1977} effectively
sets the scale of the initial vector-field perturbation
by mode mixing~\mbox{\cite{HutKlinkhamer1981,HawkingMoss1982},}
so that $|V_0(t_\text{in})| \sim T_\text{GH} \sim H_\text{dS}
                  \sim  \Lambda^{1/2}/E_\text{Planck}$
and $|d V_0/dt(t_\text{in})|\sim |H_\text{dS}\,V_0(t_\text{in})|$.
The parametric dependence of \eqref{eq:tmax} is then given by
\beqa\label{eq:T-dS-estimate}
&&
\Big[t_\text{max} - t_\text{in,\;perturbation}^\text{\,(GH-temperature)}
\Big]_\text{dS}^{(\text{vector-field\;theory})}
%%\nonumber\\[1mm]&&   %%frk
\sim
\ln\big[(E_\text{Planck})^4/\Lambda\big]\;E_\text{Planck}/\sqrt{\Lambda}\,,
\eeqa
where $t_\text{in}$ is the coordinate time of the low-frequency
(long-wavelength) matter perturbation that breaks
the original de Sitter symmetry
and where the vector-field theory considered is the one given
by \eqref{eq:model-action-Q1-EPlanck}.
It is certainly possible that a result similar to
\eqref{eq:T-dS-estimate} can be obtained for other nonstandard
matter fields, but it remains to determine precisely which types
of matter fields suffice.

For $\Lambda\to 0^{+}$ while keeping $E_\text{Planck}$ fixed,
the estimated lifetime \eqref{eq:T-dS-estimate} increases without bound,
\mbox{$[t_\text{max} - t_\text{in}]_\text{dS}$} $\to$ $+\infty$.
This behavior agrees with the naive expectation that the Minkowski
solution of the $\Lambda=0$ theory remains effectively stable,
even in the presence of the vector field $V_{\alpha}(x)$.
Note that the type of vector-field model
considered with $\Lambda>0$
can still give an infinite-age solution (with Minkowski
spacetime~\cite{EmelyanovKlinkhamer2011-CCP1-NEWTON} appearing asymptotically)
if the energy-density
function $\epsilon(Q_{1})$ in \eqref{eq:model-action} is
more complicated than the negative quadratic function \eqref{eq:epsilonQ1};
see also later comments.

The behavior found in Secs.~\ref{sec:Unstable-equilibrium}
and \ref{sec:Final-singularity} differs from that of ``normal'' matter,
which typically behaves according to the so-called cosmic-no-hair
conjecture~\cite{GibbonsHawking1977,HawkingMoss1982,Wald1983}.
Loosely speaking, the conjecture
states that, with appropriate matter content,
expanding universes that are not too irregular approach
an eternal de Sitter universe
(see also the recent paper~\cite{Yamamoto-etal2012},
which will be commented on below).
For homogenous cosmological models,
the cosmic-no-hair conjecture has been shown to
hold~\cite{Wald1983}, provided the matter obeys both
the strong energy condition (SEC) and
the dominant energy condition (DEC).
Recalling the succinct discussion of
Ref.~\cite{Visser1995} in terms of perfect fluids
(here, for simplicity,
specialized to the case of an isotropic pressure $P$),
the SEC corresponds to having $\rho+P \geq 0$ $\wedge$ $\rho+3\,P \geq 0$
and
the DEC to $\rho\geq 0$ $\wedge$ $P \in [-\rho,\,+\rho]$.

The numerical solutions with only a standard-matter
component (not shown here~\cite{endnote:rMnonzero})
agree with the expectations of the cosmic-no-hair conjecture.
But not so for the solutions
with an additional nonstandard vector-field matter component:
the model universe runs \emph{away} from the de Sitter solution
instead of towards it.
The same conclusion holds for the $r_{M}(\tau)=0$ case
presented in Fig.~\ref{fig:01}. The numerical solutions
display, in fact, a violation of the DEC on one count
($P \notin [-\rho,\,+\rho]$ for $\rho\geq 0$)
and a violation of the SEC on two counts
($\rho+P < 0$ and $\rho+3\,P < 0$).

It is, however, clear that quite reasonable physical systems may
display violations of the various
energy conditions~\cite{VisserBarcello2000},
perhaps the least surprising being the violation of the SEC.
In fact, SEC violation occurs already
for a positive gravitating vacuum energy density
($\rho_{V}=-P_{V} > 0$), which
can result from underlying microscopic
physical degrees of freedom~\cite{Volovik2009,KV2008}.
The crucial question is whether
the classical vector-field theory \eqref{eq:model-action-Q1-EPlanck}
can be made into a consistent quantum theory.
A related question is whether or not the vector-field
theory \eqref{eq:model-action-Q1-EPlanck} can be shown to arise
as an effective theory (see below for some remarks on infrared
quantum effects~\cite{Polyakov2007,Polyakov2009,KrotovPolyakov2010}).
Obviously, the interest of the present article is only mathematical
if the answer to both questions turns out to be negative.

For completeness, it should be mentioned that the inapplicability
of the cosmic-no-hair conjecture has also been discussed
recently in the context of
anisotropic inflationary models (cf. Ref.~\cite{Yamamoto-etal2012}
and references therein).
But the `mild' behavior found in these anisotropic
models~\cite{Yamamoto-etal2012} contrasts
with the `catastrophic' behavior resulting from the vector-field theory
\eqref{eq:model-action-Q1-EPlanck}, where the isotropic model universe
simply comes to an end (see Sec.~\ref{sec:Final-singularity}
and  Refs.~\cite{Caldwell1999,Starobinsky1999,McInnes2001,%
Caldwell-etal2003,Faraoni2003,Nojiri-etal2005,Dabrowski2011,Barrow-etal2012}).
Moreover, the theory considered in this
article has a genuine positive cosmological constant $\Lambda$,
not just a positive value of the
scalar potential at a particular localized field configuration,
which can make a difference for the nonlinear field equations
and certainly does make a difference for the global
spacetime structure~\cite{HawkingEllis1973}.

In fact, the global structure of de Sitter spacetime has been
argued to be responsible for a fundamental quantum instability
through particle production~\cite{Polyakov2007,Polyakov2009,KrotovPolyakov2010}.
The search is for a macroscopic
description of the corresponding backreaction effects.
Naively, our vector-field theory \eqref{eq:model-action-Q1-EPlanck}
appears to be ruled out, as the Hubble parameter
\emph{increases} due to the instability
(details in the caption of Fig.~\ref{fig:01}; see also
the middle panel of Fig.~2 in Ref.~\cite{endnote:rMnonzero}),
in a way reminiscent of what happens with an evaporating
Schwarzschild black hole. However, the same type of
vector-field theory can also give a \emph{decreasing} Hubble
parameter (see the top-right panel of Fig.~1 in
Ref.~\cite{EmelyanovKlinkhamer2011-CCP1-NEWTON}),
provided the quadratic energy-density
function $\epsilon(Q_{1})$ in \eqref{eq:model-action} is
replaced by a more complicated
function~\cite{EmelyanovKlinkhamer2011-CCP1-FRW-NEWTON,
EmelyanovKlinkhamer2011-CCP1-NEWTON,KV2008}.
Hence, if an effective vector-field theory is somehow relevant
for the macroscopic description of backreaction effects from particle
production~\cite{Polyakov2007,Polyakov2009,KrotovPolyakov2010}, then
the microscopic processes themselves select
an appropriate macroscopic $\epsilon$--type function.

\section*{%\hspace*{-4.5mm}
ACKNOWLEDGMENTS}
We thank T.Q. Do for reminding us of the cosmic-no-hair conjecture
and for bringing Ref.~\cite{Yamamoto-etal2012} to our attention.
In addition, we thank S. Thambyahpillai, G.E. Volovik, J. Weller,
and the referee
for helpful comments on an earlier version of this article.\newline
\textit{Note added.---}
Further discussion of particle-production
backreaction effects in de Sitter spacetime
can be found in Ref.~\cite{Klinkhamer2012}.


\newpage
\begin{thebibliography}{99}

\bibitem{Einstein1917}
A. Einstein,
\hspace*{0mm}``Kosmologische Betrachtungen zur allgemeinen Relativit\"{a}tstheorie,''
Sitzungsber. Preuss. Akad. Wiss. 8. Febr. 1917,
142 (1917).

\bibitem{deSitter1917}
W. de Sitter,
\hspace*{0mm}``On the relativity of inertia. Remarks concerning Einstein's latest hypothesis,''
Proc. Royal Acad. Amsterdam {\bf 19}, 1217 (1917);
\hspace*{0mm}``On the curvature of space,''
\emph{ibid.} {\bf 20}, 229 (1917).


\bibitem{HawkingEllis1973}
S.W.~Hawking and G.F.R.~Ellis,
\textit{The Large Scale Structure of Space-Time}
(Cambridge Univ. Press, Cambridge, England, 1973).

\bibitem{Weinberg1989}
S. Weinberg,
\hspace*{0mm}``The cosmological constant problem,''
Rev. Mod. Phys. {\bf 61}, 1 (1989).
%%CITATION = RMPHA,61,1;%%

\bibitem{PDG2010}
K. Nakamura {\it et al.} [Particle Data Group],
\hspace*{0mm}``Review of particle physics,''
J. Phys. G {\bf 37}, 075021 (2010), Sec. 21.
%%CITATION = JPHGB,G37,075021;%%


\bibitem{EmelyanovKlinkhamer2011-CCP1-FRW-NEWTON}
V.~Emelyanov and F.R.~Klinkhamer,
\hspace*{0mm}``Possible solution to the main cosmological constant problem,''
Phys. Rev. D {\bf 85}, 103508 (2012),
arXiv:1109.4915. %% [hep-th].
%%CITATION = ARXIV:1109.4915;%%

\bibitem{Dolgov1997}
A.D. Dolgov,
\hspace*{0mm}``Higher spin fields and the problem of cosmological constant,''
Phys. Rev.  D {\bf 55}, 5881 (1997),
arXiv:astro-ph/9608175.
%%CITATION = PHRVA,D55,5881;%%

\bibitem{EmelyanovKlinkhamer2011-CCP1-NEWTON}
V.~Emelyanov and F.R.~Klinkhamer,
\hspace*{0mm}``Reconsidering a higher-spin-field solution to the main cosmological constant problem,''
Phys. Rev. D {\bf 85}, 063522 (2012),
arXiv:1107.0961.  %%[hep-th]
%%CITATION = ARXIV:1107.0961;%%

\bibitem{Polyakov2007}
A.M. Polyakov,
\hspace*{0mm}``De Sitter space and eternity,''
Nucl. Phys. B {\bf 797}, 199 (2008),
arXiv:0709.2899.  %% [hep-th]].
%%CITATION = ARXIV:0709.2899;%%

\bibitem{Polyakov2009}
A.M. Polyakov,
\hspace*{0mm}``Decay of vacuum energy,''
Nucl. Phys. B {\bf 834}, 316 (2010),
arXiv:0912.5503. %% [hep-th]].
%%CITATION = ARXIV:0912.5503;%%

\bibitem{KrotovPolyakov2010}
D. Krotov and A.M. Polyakov,
\hspace*{0mm}``Infrared sensitivity of unstable vacua,''
Nucl. Phys. B {\bf 849}, 410 (2011),
arXiv:1012.2107. %% [hep-th]].
%%CITATION = ARXIV:1012.2107;%%

\bibitem{Weinberg1964}
S. Weinberg,
\hspace*{0mm}``Derivation of gauge invariance and the equivalence principle from Lorentz invariance of the S--matrix,''
Phys. Lett. {\bf 9}, 357 (1964).

\bibitem{Verhulst1996}
F. Verhulst,
\textit{Nonlinear Differential Equations and Dynamical Systems},
Second Edition
(Springer, Berlin, 1996).

\bibitem{endnote:rMnonzero}
Numerical results for the $r_{M}\ne 0$ case are shown in
the first three figures of an earlier preprint version
of the present article,
V.~Emelyanov and F.R.~Klinkhamer,
arXiv:1204.5085v5.

\bibitem{Caldwell1999}
R.R.~Caldwell,
\hspace*{0mm}``A phantom menace?,''
Phys.\ Lett.\ B {\bf 545}, 23 (2002),
arXiv:astro-ph/9908168.
%%CITATION = ASTRO-PH/9908168;%%

\bibitem{Starobinsky1999}
A.A.~Starobinsky,
\hspace*{0mm}``Future and origin of our universe: Modern view,''
Gravitation Cosmol.  {\bf 6}, 157 (2000),
arXiv:astro-ph/9912054.
%%CITATION = ASTRO-PH/9912054;%%

\bibitem{McInnes2001}
B.~McInnes,
\hspace*{0mm}``The dS/CFT correspondence and the big smash,''
JHEP {\bf 0208}, 029 (2002),
arXiv:hep-th/0112066.
%%CITATION = HEP-TH/0112066;%%

\bibitem{Caldwell-etal2003}
R.R.~Caldwell, M.~Kamionkowski, and N.N.~Weinberg,
\hspace*{0mm}``Phantom energy and cosmic doomsday,''
Phys.\ Rev.\ Lett.\  {\bf 91}, 071301 (2003)
arXiv:astro-ph/0302506.
%%CITATION = ASTRO-PH/0302506;%%

\bibitem{Faraoni2003}
V.~Faraoni,
\hspace*{0mm}``Possible end of the universe in a finite future from dark energy with  $w<-1$,''
Phys. Rev. D {\bf 68}, 063508 (2003),
arXiv:gr-qc/0307086.
%%CITATION = GR-QC/0307086;%%

\bibitem{Nojiri-etal2005}
S.~Nojiri, S.D.~Odintsov, and S.~Tsujikawa,
\hspace*{0mm}``Properties of singularities in (phantom) dark energy universe,''
Phys.\ Rev.\ D {\bf 71}, 063004 (2005),
arXiv:hep-th/0501025.
%%CITATION = HEP-TH/0501025;%%

\bibitem{Dabrowski2011}
M.P.~Dabrowski,
\hspace*{0mm}``Spacetime averaging of exotic singularity universes,''
Phys.\ Lett.\ B {\bf 702}, 320 (2011),
arXiv:1105.3607.  %% [gr-qc]].
%%CITATION = ARXIV:1105.3607;%%

\bibitem{Barrow-etal2012}
J.D.~Barrow, A.B.~Batista, J.C.~Fabris, M.J.S.~Houndjo, and G.~Dito,
\hspace*{0mm}``Quantum effects near future singularities,''
arXiv:1201.1138. %% [gr-qc].
%%CITATION = ARXIV:1201.1138;%%

\bibitem{GibbonsHawking1977}
G.W.~Gibbons and S.W.~Hawking,
\hspace*{0mm}``Cosmological event horizons, thermodynamics, and particle creation,''
Phys. Rev. D {\bf 15}, 2738 (1977).
%%CITATION = PHRVA,D15,2738;%%

\bibitem{HutKlinkhamer1981}
P.~Hut and F.R.~Klinkhamer,
\hspace*{0mm}``Global space-time effects on first-order phase transitions from grand unification,''
Phys. Lett. B {\bf 104}, 439 (1981).
%%CITATION = PHLTA,B104,439;%%


\bibitem{HawkingMoss1982}
S.W.~Hawking and I.G.~Moss,
\hspace*{0mm}``Supercooled phase transitions in the very early universe,''
Phys. Lett. B {\bf 110}, 35 (1982).
%%CITATION = PHLTA,B110,35;%%

\bibitem{Wald1983}
R.M.~Wald,
\hspace*{0mm}``Asymptotic behavior of homogeneous cosmological models in the presence of a positive cosmological constant,''
Phys. Rev. D {\bf 28}, 2118 (1983).
%%CITATION = PHRVA,D28,2118;%%

\bibitem{Yamamoto-etal2012}
K.~Yamamoto, M.~Watanabe, and J.~Soda,
\hspace*{0mm}``Inflation with multi-vector-hair: The fate of anisotropy,''
Class. Quant. Grav.  {\bf 29}, 145008 (2012),
arXiv:1201.5309.  %% [hep-th].
%%CITATION = ARXIV:1201.5309;%%

\bibitem{Visser1995}
M.~Visser,
\textit{Lorentzian wormholes: From Einstein to Hawking}
(AIP, Woodbury, USA, 1995), Chap. 12.

\bibitem{VisserBarcello2000}
M.~Visser and C.~Barcelo,
\hspace*{0mm}``Energy conditions and their cosmological implications,''
arXiv:gr-qc/0001099.
%%CITATION = GR-QC/0001099;%%

\bibitem{Volovik2009}
G.E.~Volovik,
\textit{The Universe in a Helium Droplet}
(Oxford Univ. Press, Oxford, England, 2009), Secs. 3.2 and 7.3.
%%CITATION = IMPHA,117,1;%%

\bibitem{KV2008}
F.R. Klinkhamer and G.E. Volovik,
\hspace*{0mm}``Self-tuning vacuum variable and cosmological constant,''
Phys. Rev. D {\bf 77}, 085015 (2008),
arXiv:0711.3170.
%%CITATION = PHRVA,D77,085015;%%

\bibitem{Klinkhamer2012}
F.R.~Klinkhamer,
\hspace*{0mm}``On vacuum-energy decay from particle production,''
arXiv:1205.7072.  %% [hep-th].
%%CITATION = ARXIV:1205.7072;%%

%arXiv:1205.7072

\end{thebibliography}
\end{document}